\documentclass{juliacon}
\newcommand{\ts}{\textsuperscript}
\hypersetup{hidelinks} % hide ugly link boxes

\begin{document}

% **************GENERATED FILE, DO NOT EDIT**************

\title{High-performance xPU Stencil Computations in Julia}

\author[1]{Samuel Omlin}
\author[2, 3]{Ludovic R\"ass}
\affil[1]{Swiss National Supercomputing Centre (CSCS), ETH Zurich, Lugano, Switzerland}
\affil[2]{Laboratory of Hydraulics, Hydrology and Glaciology (VAW), ETH Zurich, Zurich, Switzerland}
\affil[3]{Swiss Federal Institute for Forest, Snow and Landscape Research (WSL), Birmensdorf, Switzerland}


\hypersetup{
pdftitle = {High-performance xPU Stencil Computations in Julia},
pdfsubject = {JuliaCon 2019 Proceedings},
pdfauthor = {Samuel Omlin, Ludovic R\"ass},
pdfkeywords = {Julia, xPU, GPU, Stencil Computations, Code Generation, Architecture-agnostic, Shared Memory Parallelization, Communication-Computation Overlap, Supercomputing},
}

\maketitle

\begin{abstract}

We present an efficient approach for writing architecture-agnostic parallel high-performance stencil computations in Julia, which is instantiated in the package \texttt{ParallelStencil.jl}. Powerful metaprogramming, costless abstractions and multiple dispatch enable writing a single code that is suitable for both productive prototyping on a single CPU thread and production runs on multi-GPU or CPU workstations or supercomputers. We demonstrate performance close to the theoretical upper bound on GPUs for a 3-D heat diffusion solver, which is a massive improvement over reachable performance with \texttt{CUDA.jl} Array programming.

\end{abstract}

\section{Introduction}
Graphics processing units (GPUs) capable of general-purpose computing have revolutionized the hardware industry and as a result High Performance Computing (HPC) since the dawn of the 21\ts{st} century. While industry and academia are doing their best to adapt their software to the new hardware landscape, the latter continues to be reshaped constantly. In addition, new unconventional highly innovative hardware developments driven by the powerful AI industry (e.g. the Cerebras WSEs and Graphcore IPUs) draw up yet the next potential hardware revolution. In the light of the high pace and increasing diversity in hardware evolution, the HPC community has identified the 3 ``P''s - (scalable) Performance, (performance) Portability and Productivity - as fundamental requirements for today's and tomorrow's software development. The approach and package development presented in this paper responds to each of the 3 ``P''s. We present an approach for automatic parallelization and optimization of architecture-agnostic stencil computations deployable on both GPU and CPU (in the remainder we use xPU to refer simultaneously to GPU and CPU); the computations can furthermore automatically hide the communication needed for distributed parallelization as required for large scale supercomputing.

\begin{figure}[t]
\begin{lstlisting}[language = Julia, numbers=left, numberstyle=\tiny\color{gray}]
using ParallelStencil
using ParallelStencil.FiniteDifferences3D
@init_parallel_stencil(CUDA, Float64, 3)

@parallel loopopt=true optvars=T function step!(
    T2, T, Ci, lam, dt, _dx, _dy, _dz)
    @inn(T2) = @inn(T) + dt*(
        lam*@inn(Ci)*(@d2_xi(T)*_dx^2 + 
                      @d2_yi(T)*_dy^2 + 
                      @d2_zi(T)*_dz^2 ) )
    return
end

function diffusion3D()
    # Physics
    lam      = 1.0           #Thermal conductivity
    c0       = 2.0           #Heat capacity
    lx=ly=lz = 1.0           #Domain length x|y|z

    # Numerics
    nx=ny=nz = 512           #Nb gridpoints x|y|z
    nt       = 100           #Nb time steps
    dx       = lx/(nx-1)     #Space step in x
    dy       = ly/(ny-1)     #Space step in y
    dz       = lz/(nz-1)     #Space step in z
    _dx, _dy, _dz = 1.0/dx, 1.0/dy, 1.0/dz

    # Initial conditions
    T  = @ones(nx,ny,nz).*1.7 #Temperature
    T2 = copy(T)              #Temperature (2nd)
    Ci = @ones(nx,ny,nz)./c0  #1/Heat capacity

    # Time loop
    dt = min(dx^2,dy^2,dz^2)/lam/maximum(Ci)/6.1
    for it = 1:nt
        @parallel loopopt=true step!(
            T2, T, Ci, lam, dt, _dx, _dy, _dz)
        T, T2 = T2, T
    end

end

diffusion3D()

\end{lstlisting}

\caption{Stencil-based 3-D heat diffusion xPU solver implemented using ParallelStencil with time step kernel written in math-close notation.}
	\label{fig:code}
\end{figure}

\section{Approach}
Our approach for the expression of architecture-agnostic high-performance stencil computations relies on the usage of Julia's powerful metaprogramming capacities, costless high-level abstractions and multiple dispatch. We have instantiated the approach in the Julia package \texttt{ParallelStencil.jl}. Using ParallelStencil, a simple call to the macro \texttt{@parallel} is sufficient to parallelize and launch a kernel that contains stencil computations, which can be expressed explicitly or with math-close notation. Fig.~\ref{fig:code} shows a stencil-based 3-D heat diffusion xPU solver implemented using ParallelStencil, where the kernel defining an explicit time step is written in math-close notation (lines 5-12) and the macro \texttt{@parallel} is used for its parallelization (line 5) and launch (line 36). 

The package used underneath for parallelization is defined in a initialization call beforehand (Fig.~\ref{fig:code}, line 3). Currently supported are \texttt{CUDA.jl} \cite{besard2018effective} for running on GPU and \texttt{Base.Threads} for CPU. Leveraging metaprogramming, ParallelStencil automatically generates high-performance code suitable for the target hardware, and automatically derives kernel launch parameters from the kernel arguments by analyzing the bounds of the contained arrays. Certain stencil-computation-specific optimizations leveraging, e.g., the on-chip memory of GPUs need to be activated with keyword arguments to the macro \texttt{@parallel} (Fig. \ref{fig:code}, line 5). A set of architecture-agnostic low level kernel language constructs allows for explicit low level kernel programming when useful, e.g., for the explicit control of shared memory on the GPU (these low level constructs are GPU-computing-biased). 

\begin{figure}[t]
    \centerline{\includegraphics[width=8cm]{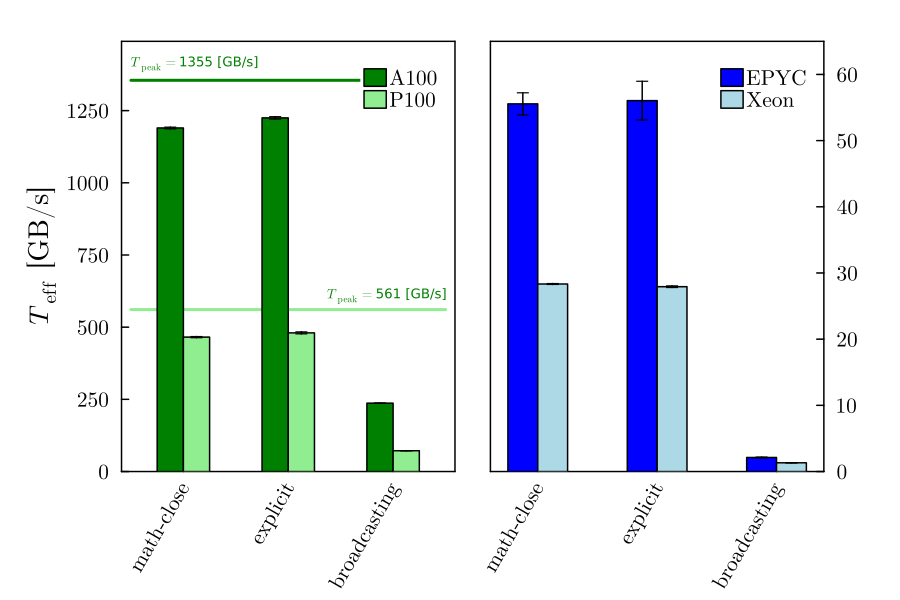}}
    \caption{Effective memory throughput $T_\mathrm{eff}$ for Nvidia Tesla A100 SXM4 and P100 PCIe GPUs, respectively, and for 2 x 16 Core AMD EPYC 7282 and 12 Core Intel Xeon E5-2690 v3 CPUs, respectively. The error bars visualize the 95\% confidence interval of the reported medians (20 samples). The raw data and plotting script are available in \url{github.com/omlins/ParallelStencil.jl/tree/JuliaConProceeding2022/paper}.}
	\label{fig:performance}
\end{figure}

Arrays are automatically allocated on the hardware chosen for the computations (GPU or CPU) when using the allocation macros provided by ParallelStencil (Fig.~\ref{fig:code}, lines 29-31), avoiding any need of code duplication. Moreover, the allocation macros are fully declarative in order to let ParallelStencil choose the best data layout in memory. Notably, logical arrays of structs (or of small arrays) can be either laid out in memory as arrays of structs or as structs of arrays accounting for the fact that each of these allocation approaches has its use cases where it performs best.

ParallelStencil is seamlessly interoperable with packages for distributed parallelization, as e.g. \texttt{ImplicitGlobalGrid.jl} \cite{implicitglobalgrid2022} or \texttt{MPI.jl}, in order to enable high-performance stencil computations on GPU or CPU supercomputers. Communication can be hidden behind computation with as simple macro call \cite{implicitglobalgrid2022}. The usage of this feature solely requires that communication can be triggered explicitly as it is possible with, e.g., ImplicitGlobalGrid and \texttt{MPI.jl}.

\section{Results}
We here report the performance achieved on different architectures with the 3-D heat diffusion xPU solver (Fig.~\ref{fig:code}) and of an equivalent solver with explicit notation for the stencil computations and compare it to the performance obtained with a Julia solver written in a traditional way using GPU or CPU array broadcasting. We observe that using ParallelStencil we achieve an effective memory throughput, $T_\mathrm{eff}$, of 496~GB/s and 1262~GB/s on the Nvidia P100 and A100 GPUs, which can reach a peak throughput, $T_\mathrm{peak}$, of 561~GB/s and 1355~GB/s, respectively; this means we reach 88\% and 93\% of the respective hardware's theoretical performance upper bound ($T_\mathrm{eff}$ and its interpretation are explained, e.g., in \cite{rass2022assessing}). Furthermore, using ParallelStencil we obtain a speedup of up to a factor $\approx 5$ and 
$\approx 29$ over the versions with GPU and CPU array broadcasting (the latter is not capable of multi-threading), respectively.
Moreover, we have translated solvers for highly nonlinear 3-D poro-visco-elastic two-phase flow and 3-D reactive porosity waves written in CUDA C using MPI to Julia by employing ParallelStencil and ImplicitGlobalGrid and compared obtained performance. The translated solvers achieved 90\% and 98\% of the performance of the respective original CUDA C solvers. In addition, relying on ParallelStencil`s feature to hide communication behind computation, the 3-D poro-visco-elastic two-phase flow solver achieved over 95\% parallel effiency on up to 1024 GPUs \cite{implicitglobalgrid2022}. 

\section{Conclusions}
We have shown that ParallelStencil enables scalable performance, performance portability and productivity and responds to the challenge of addressing the 3 ``P''s in all of its aspects. Moreover, we have outlined the effectiveness and wide applicability of our approach within geosciences. Our approach is naturally in no sense limited to geosciences as stencil computations are commonly used in many disciplines across all of science. We illustrated this in recent contributions, where we showcased a computational cognitive neuroscience application modelling visual target selection using ParallelStencil and \texttt{MPI.jl} \cite{pasc22} and a quantum fluid dynamics solver using the nonlinear Gross-Pitaevski equation implemented with ParallelStencil and ImplicitGlobalGrid \cite{pasc21}.

\section{Acknowledgments}
 This work was supported by a grant from the Swiss National Supercomputing Centre (CSCS) under project ID c23 through the Platform for Advanced Scientific Computing (PASC) program. We acknowledge A100 DGX-1 computing resources at VAW, ETH Zurich.

% **************GENERATED FILE, DO NOT EDIT**************

\bibliographystyle{juliacon}
\bibliography{ref.bib}

\end{document}